# Effects of Mixed Distribution Statistical Flood Frequency Models on Dam Safety Assessments: A Case Study of the Pueblo Dam, USA

K. J. Roop-Eckart[1], Sanjib Sharma[2, *], Mahkameh Zarekarizi[3], Ben Seiyon Lee[4], Caitlin Spence[5], Tess Russo[1,6], and Klaus Keller[1,2,7]

[1]Department of Geosciences, Pennsylvania State University, University Park, PA 16802, USA
[2]Earth and Environmental Systems Institute, Pennsylvania State University, University Park, PA, USA
[3]Jupiter Intelligence, San Mateo, CA 94401, USA
[4]Department of Statistics, The George Mason University, Fairfax, VA, USA
[5]Data Services, Metropolitan Area Planning Council, Boston MA 02111, USA
[6]Department of Mathematics, Pennsylvania State University, University Park, PA 16802, USA
[7]Thayer School of Engineering, Dartmouth College, Hanover, NH 03755, USA

Corresponding Author: Sanjib Sharma (sanjibsharma66@gmail.com)

**Abstract**
Statistical flood frequency analysis coupled with hydrograph scaling is commonly used to generate design floods to assess dam safety assessment. The safety assessments can be highly sensitive to the choice of the statistical flood frequency model. Standard dam safety assessments are typically based on a single distribution model of flood frequency, often the Log Pearson Type III or Generalized Extreme Value distributions. Floods, however, may result from multiple physical processes such as rain on snow, snowmelt or rainstorms. This can result in a mixed distribution of annual peak flows, according to the cause of each flood. Engineering design choices based on a single distribution statistical model are vulnerable to the effects of this potential structural model error. To explore the practicality and potential value of implementing mixed distribution statistical models in engineering design, we compare the goodness of fit of several single- and mixed-distribution peak flow models, as well as the contingent dam safety assessment at Pueblo, Colorado as a didactic example. Summer snowmelt and intense summer rainstorms are both key drivers of annual peak flow at Pueblo. We analyze the potential implications for the annual probability of overtopping-induced failure of the Pueblo Dam as a didactic example. We address the temporal and physical cause separation problems by building on previous work with mixed distributions. We find a Mixed Generalized Extreme Value distribution model best fits peak flows observed in the gaged record, historical floods, and paleo floods at Pueblo. Finally, we show that accounting for mixed distributions in the safety assessment at Pueblo Dam increases the assessed risk of overtopping.

# 1. Introduction
## 1.1 Dam safety assessment in the United States

People rely on dams for flood protection, and catastrophic dam failure can be devastating [1,2]. For example, the Johnstown Dam failure in 1977 killed 40 people and caused millions of dollars in damages, and the Teton Dam failure in 1976 killed 11 people and caused hundreds of millions of dollars in damages [3]. One of the primary modes of dam failure is overtopping, i.e., peak reservoir water stage exceeds the dam height [4]. While this failure mode is most common in earthen embankment dams, it is a major cause of failure in all dam designs [4]. In the United States, the Bureau of Reclamation (USBR) oversees the safety and long-term maintenance of many of the flood control dams in the western United States. The USBR uses several methods for determining dam overtopping probability based on the available data and the safety needs of the dam.

First, can the dam safely route the largest reasonably plausible discharge, given known meteorological and hydrological mechanisms and their limits in the basin? The probable maximum precipitation (PMP) is the maximum precipitation event that can be reasonably expected to occur given the current understanding of meteorological factors, and is derived from a meteorological assessment of the region [6]. The probable maximum flood (PMF) represents the largest runoff scenario that can be reasonably expected to occur given the PMP [5]. If a dam can route the PMF without overtopping, the dam is considered safe in all reasonably possible flood scenarios [5,7].

Second, if the PMF would cause the dam to be overtopped, then methods to estimate the overtopping return period (i.e., average recurrence interval) are employed to complete the safety assessment [5]. One relatively quick and low-cost approach to estimating the overtopping return period is statistical flood frequency analysis with hydrograph scaling [5]. A statistical distribution is fitted to a dataset of annual peak flows, historical flood peak flows, and paleo flood peak flow bounds to create a flood frequency curve for peak flows. A representative flood hydrograph is then scaled to peak flows from the flood frequency curve. The flood hydrographs are then used in routing simulations for the dam to determine the probability of overtopping [5,7]. If the probability of overtopping is less than the acceptable risk of loss of life, the dam is considered safe.

## 1.2 Statistical flood frequency analysis

Traditional flood frequency analysis creates flood frequency curves by fitting a statistical distribution to observed flood peak magnitude data. Flood frequency curves can be used to extrapolate peak flows at return periods exceeding the time spanned by the observational record, such as the design return period for a dam. However, dam safety depends on both the peak flow and the volume delivered over time. Hydrologic hazard curves relate peak flow and volume for specified durations to return periods to inform dam-safety assessments [5]. To produce hydrologic hazard curves for dam safety assessments, peak flows are used to linearly scale the magnitude of representative flood hydrographs [5]. The scaled flood hydrographs are then used as inputs to a routing model to determine the peak reservoir stage. This method is called hydrograph scaling [5]. The combination of statistical flood frequency analysis and hydrograph scaling represents a reasonably fast and efficient method to determine the return period of dam overtopping flood events, and thus the safety of dams [5].

Statistical flood frequency analysis typically operates under the assumption that annual peak discharges, historical floods magnitudes, and paleo flood bounds can be described with a single distribution [8]. However, in some cases, a mixed distribution of peak flows can occur due to multiple physical flood mechanisms [8–10]. There are several procedures for identifying mixed distributions in flood frequency analysis [8]. The current federal guidelines for statistical flood frequency analysis in the United States, Bulletin 17C [8], requires prior knowledge about the specific causes of each observed peak flow event. The data is separated based on physical cause, and distributions are fitted to the peak discharge data for each individual flood mechanism [8]. The multiple distributions are then combined into a composite distribution [11]. However, in some cases, this prior knowledge of physical cause may not be available due to sparse data records or a lack of additional information about each peak discharge event. There is a need for further work on the identification and treatment of mixed distribution flood flows in flood frequency analysis [8,12].

## 1.3 Conflicting assessments at the Pueblo Dam, CO, USA

The assessed safety of the Pueblo Dam has come under scrutiny since its construction in the 1970s [6,8,10]. The dam was designed to safely route the PMF estimated at the time of its construction. However, a revised PMF, calculated based on updated watershed and extreme storm characteristics, would overtop the dam [6]. Thus, the dam's probability of overtopping needed to be established to determine whether the dam met the United States

Bureau of Reclamation safety regulations. The United States Bureau of Reclamation requires all its infrastructure to be safe to one life lost per 1,000 years of service [13]. A loss of life study was conducted on the Pueblo Dam and determined that between 131 and 376 people would die from catastrophic failure due to overtopping depending on whether failure occurred at night or during the day [13]. Therefore, the overtopping return period for the Pueblo Dam must be greater than 131,000 to 376,000 years to meet USBR safety standards. Safety analyses of the dam typically round this return period to 400,000 years [13].

A flood frequency assessment examined the return periods of peak design inflows and outflows [10]. England et al. (2010)[10] hypothesized the presence of a mixed distribution of annual peak flows caused by both snowmelt and rainfall, however, did not address the effects a mixed distribution may have on the flood frequency analysis. Current methods require that mixed distributions be separable by individual cause, or by mechanisms that are separable by season [8]. Statistical flood frequency analysis at the Pueblo Dam is complicated by both mechanisms occurring at the same time of year [10]. Additionally, many of the peak flows are not explicitly attributed to a particular physical mechanism, i.e., rainfall or snowmelt [10].

England et al. (2014)[14] uses a physically-based model to assess the safety of the Pueblo Dam by simulating watershed response to extreme rainfall storms, which are assumed to be drivers of the most extreme peak flows. England et al. (2014) [14] concludes the Pueblo Dam overtopping return period meets Bureau of Reclamation safety standards. Safety assessments using physically-based rainfall-runoff models and stochastic storm transposition typically require more location-specific input data and computational time [5,10,14] and are subject to uncertainties in watershed discharge routing and storm characteristics [14].

### 1.4 Study design and focus

We assess mixed distribution statistical flood frequency models' ability to model peak flows hypothesized to be generated by multiple mechanisms. We use goodness-of-fit criteria to evaluate the ability of these models to identify the mixed distributions. In doing so, we address three main research questions:
1. Is there a statistically identifiable mixed distribution of peak discharges at Pueblo?
2. Which distribution is most likely given the data?
3. Does the best fitting model change the dam safety assessment?

This study builds on prior work in likelihood functions [15,16] and mixed distributions [9,17]. This study addresses both identification and treatment of mixed distributions at a study area where current methods of separation by physical cause are not adequate due to sparse data. The study uses mixed distributions and robust goodness-of-fit criteria to statistically establish the presence of a mixed distribution of annual peak flows and perform flood frequency analysis of the mixed distribution. The didactic safety assessment illustrates the importance of considering the physical origins of the data used in statistical models and accounting for model uncertainty.

## 2 Methods
### 2.1 Site description and data sources

A map of Arkansas River Watershed, just to the upstream of Pueblo dam is shown in Figure 1. Annual peak flows at Pueblo, Colorado, consisting of two distributions caused by two physical processes (Fig. 1, Fig. 2) [10]. The summer snowmelt peak flows are often smaller, with peak flows less than 283 $m^3s^{-1}$, while the peak flows caused by summer rainstorms are often larger than 283 $m^3s^{-1}$ (Fig. 1, Fig. 2) [10].

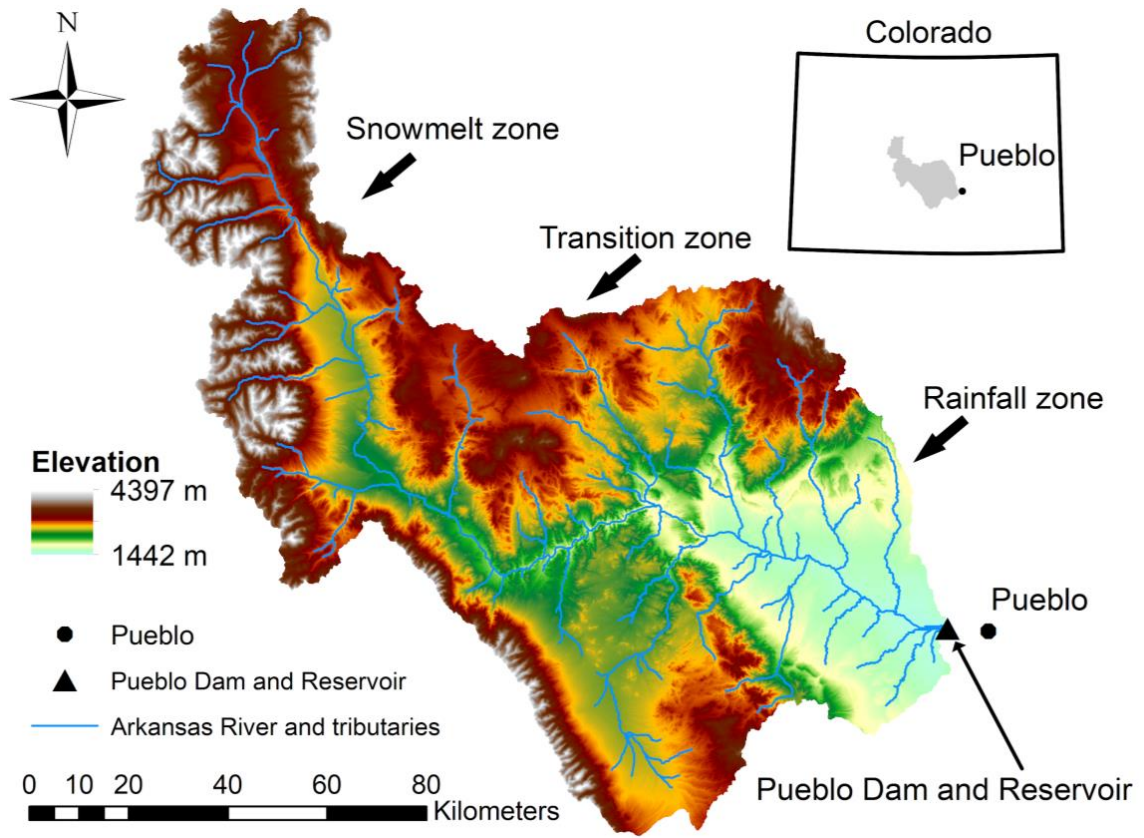

**Fig 1.** Map of the upper Arkansas River watershed, upstream of Pueblo, CO. The watershed can be divided into three zones according to the dominant physical mechanism of the largest floods generated within each zone. In the mountain headwaters of the Arkansas River, summer snowmelt is the dominant cause of peak flows. At transitional elevations, summer rainstorms and snowmelt can both contribute to the largest peak flows. Where the narrow valley opens into the plains, snowmelt floods may still be observed, but the largest peak flows are caused by extreme rainfall events [10]. Elevations are from the National Elevation Dataset [18]; the state boundary of Colorado is from the United States Census Bureau; and the Arkansas River and tributaries were delineated from the elevation data in ArcMap 10.3.1.

We compile published stream gage, historical flood, and paleoflood data at Pueblo, Colorado. The peak discharge data for Pueblo, Colorado, (Fig. 2) consists of an 81-year annual peak flow record, three historical floods, and one paleoflood bound. The 81-year gage record is from USGS gage 07099500 at Pueblo, with records from 1895 to 1975. The peak flows of the three historical floods in 1864, 1893, and 1894 and paleoflood bound are calculated and reported in Campbell (1922) [19], Follansbee and Jones (1922) [20], Baker and Hafen (1927)[21], Hafen (1948) [22], and England et al. (2010) [10] based on historical records of the peak flood elevations.

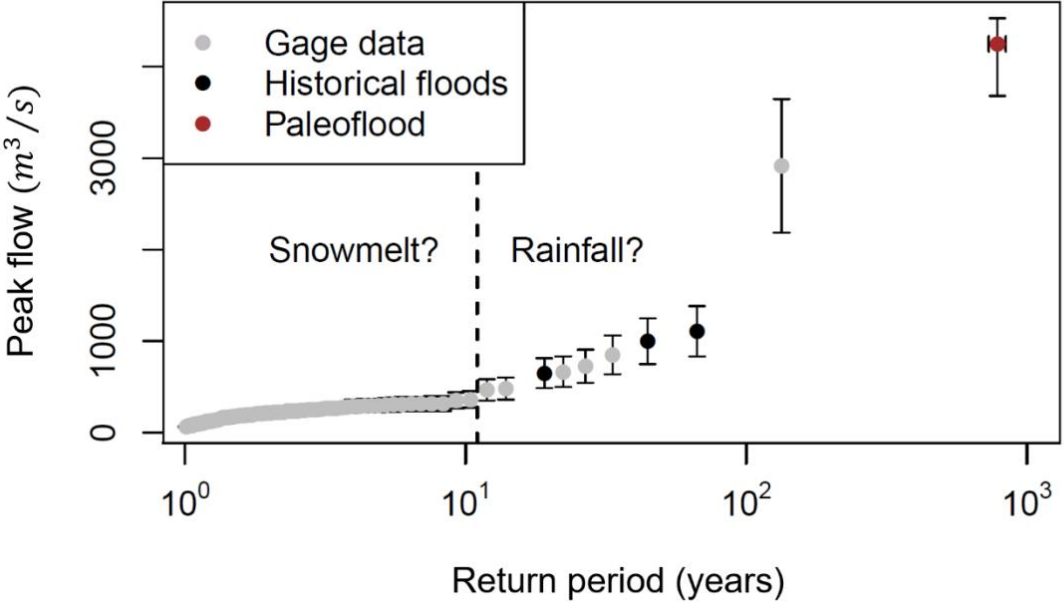

**Fig 2.** Return period plot of annual peak flows, historical floods [10,19–22], and the paleoflood [10] at Pueblo, Colorado. The dashed line represents a transition at the 11-year return period (or 283 $m^3s^{-1}$) delineating the two different causes of peak discharge events [10]. The error bars represent the 95% confidence interval for flood magnitude and timing observations, estimated from generalized measurement error estimates in O'Connell et al. (2002) [15]. The error bounds on the paleoflood magnitude and return period are calculated in England et al. (2010) [10] using the HEC-RAS hydraulic flow model on seven-channel cross-sections.

We were unable to find complete uncertainty estimates for the USGS stream gage annual peak flows and for the historical floods. Additionally, while the paleoflood bound is estimated with upper and lower bounds [10], a probability distribution is not assigned to the estimated error. We estimate uncertainties for gaged annual peak flows and historical floods with uncertainty estimates for western United States stream gages and historical floods based on assumed normally distributed errors for annual peak flows and historical floods and triangular probability distribution paleoflood bounds based on the upper and lower bound discharge estimates and paleoflood age bound uncertainty distributions [15].

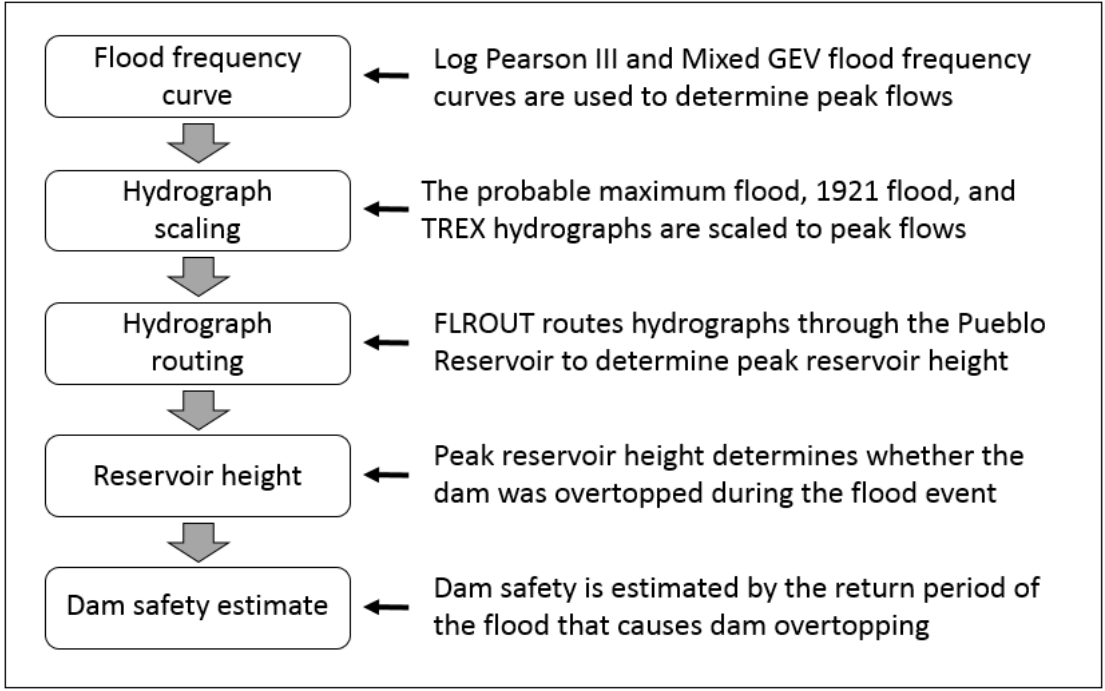

**Fig 3.** Flow chart illustrating the overall workflow of the methods in this analysis. FLROUT- the Bureau of Reclamation flood routing program [13].

**2.2 Likelihood functions**

We use maximum likelihood estimates of statistical flood frequency model parameters based on annual peak flow stream gage records, historical flood records, and paleoflood data at Pueblo. We adapt the standard likelihood formula (Eq. 1) to account for differences in timescales of annual, historical, and paleo observations (Eq. 2) [16], and for peak flow observation errors (Eq. 3, Eq. 4). Additionally, we can account for observation errors in the age of paleoflood events (Eq. 5).

The likelihood function for model parameters based on continuous annual peak flow data from the gage record is

$$L(\theta|x) = \prod f(x_i), \quad (1)$$

where $L()$ is the likelihood, $\theta$ is the parameter set of the distribution, $x$ is the annual peak flow magnitude, and $f()$ is the probability density function (pdf).

The likelihood function for model parameters based on censored data or data which is only partially observed, such as historical or paleo floods incorporates a binomial distribution to account for threshold censoring and normalizes to the total probability mass of the flood frequency distribution above the peak flow threshold value. The likelihood function for censored data is then

$$L(\theta|y) = \left\{ \binom{h}{k} F(X_0)^{h-k}[1-F(X_0)]^k \right\} \prod f(y_i)/[1-F(X_0)], \quad (2)$$

where $F()$ is the cumulative density function (cdf), $f()$ is the pdf, $X_0$ is the peak flow threshold exceedance value, $h$ is the length of the historical paleo flood record in years, $k$ is the number of threshold exceedances, and $y$ are the values of the threshold exceedances (i.e. recorded historical or paleo floods). The binomial distribution and cdf assess the probability of the peak flow exceeding the peak flow threshold, while the pdf assesses the probability of the peak flow given the distribution.

We adapt the likelihood functions to incorporate peak flow measurement uncertainty based on the methods by O'Connell et al. (2002)[15]. The likelihood function for annual peak flow gage data with uncertainty calculates the probability density for several possible values of each data point and multiplies the probability density by the values' probability of being the true value given an estimated distribution of the peak flow measurement error. The likelihood function is then

$$L(\theta|x) = \prod_{i=1}^{s}\left[\sum_j f_{ij}^s f(x_{ij})\right] \quad (3)$$

where $L()$ is the likelihood, $\theta$ is the parameter set of the distribution, $x$ is the annual maximum discharge, s is the length of the data, $x_{ij}$ is a data point $x_i$ with an error from the Gaussian error distribution, and $f^s_{ij}$ is a discrete pdf of $x_{ij}$ given the error distribution, $i$ is the $i$th annual peak flow in the gage record, and j is the $j$th draw from the measurement error. Each discrete probability in $f^s_{ij}$ corresponds to a peak flow value $x_{ij}$ from the Gaussian error distribution.

The likelihood function for historical floods and paleofloods follows the same formula as Eq. (3). Thus, the likelihood function is

$$L(\theta|y) = \prod_{i=1}^{v}\left[\sum_j f_{ij}^y L(\theta|y_{ij})\right], \quad (4)$$

where $y$ is the historical flood or paleoflood data, $v$ is the length of the data, $y_{ij}$ is a data point $y_i$ plus an error from the Gaussian error distribution, and $f^s_{ij}$ is a discrete pdf of $y_{ij}$ given the error distribution. Substitute Eq. (4) for the likelihood in Eq. (2), to account for both threshold exceedance and measurement error in historical flood and paleoflood data.

Maximum likelihood parameter estimation with paleoflood data involves both age and peak flow uncertainty. Paleoflood age and peak flow uncertainty may be accounted for simultaneously using the log-likelihood function

$$\ln[L(\theta|T)] = \sum_{i=1}^{t}\left(\sum_j f_{ij}^{tn} n_{ij}\right) \ln\left[\sum_k f_{ik}^{td} \int_{X_{min}}^{tik} L(\theta|y)\right], \quad (5)$$

where $T$ is the paleo event age data, $t$ is the number of paleoflood event observations, $f^n_{ij}$ is the discrete probability of an age $n_{ij}$, $n_{ij}$ is the age of a paleoflood event $n_i$ plus an error from an error distribution, $f^{td}_{ik}$ is the discrete pdf of the paleoflood bounds, $t^{ik}$ are the ranges of the upper y limits for the paleoflood bounds, and $y_{min}$ is the threshold exceedance value that dictates the minimum observable paleoflood bound [15].

### 2.3 Statistical Distributions

We consider three single distributions common in flood frequency analysis: The log normal (LN2), the Log Pearson III (LP3), and the Generalized Extreme Value (GEV) distributions. The LN2 has a long history in flood frequency analysis [23]. The LP3 is the current government standard in the United States for flood frequency analysis [8], while the GEV is popular in hydroclimatic applications [24].

The probability density function of the LN2 distribution is given by

$$f(x|\mu, \sigma) = \frac{1}{x\sigma\sqrt{2\pi}} e^{-(\ln(x)-\mu)^2/2\sigma^2} \qquad (6)$$

where $f()$ is the probability density function, $x$ is the data, $\mu$ is the distribution log mean, and $\sigma$ is the log standard deviation.

The probability density function of the LP3 distribution from Bulletin 17C [8] is given by

$$f(x|\tau, \alpha, \beta) = \frac{\left(\frac{x-\tau}{\beta}\right)^{\alpha-1} e^{\left(-\frac{x-\tau}{\beta}\right)}}{|\beta|\Gamma(\alpha)}, \qquad (7)$$

where $f()$ is the probability density function, $x$ is the data, $\tau$ is the location parameter, and $\alpha$ is the shape parameter, and $\beta$ is the scale parameter.

The probability density function of the GEV distribution is given by

$$f(x|\mu, \sigma, \zeta) = \left[\frac{1}{\sigma} t(x)^{\zeta+1} e^{-t(x)}\right], \qquad (8)$$

when $t()$ is

$$t(x) = \left(1 + \zeta\left(\frac{x-\mu}{\sigma}\right)\right)^{\frac{-1}{\zeta}} \quad \text{if } \zeta \neq 0$$
$$t(x) = e^{-\frac{x-\mu}{\sigma}} \quad \text{if } \zeta = 0,$$

where $f()$ is the probability density function, $x$ is the data, $\mu$ is the location parameter, $\sigma$ is the scale parameter, and $\zeta$ is the shape parameter. We use the GEV equations as implemented in the fExtremes R package [25].

In addition to the single distributions, we consider three mixed distributions, the Two-Component Extreme Value, the Mixed (Two-Population) Generalized Extreme Value, and the Mixed Log Pearson III distributions (Table 1). The Two-Component Extreme Value distribution, derived from a compound Poisson process, was popularized for flood frequency analysis when Rossi et al. (1984)[9] applied it to 39 annual peak flows that struggled with outliers. Rossi et al. (1984) [9]assumes that these outliers are the product of a second, upper flood distribution.

The probability density function of the Two-Component Extreme Value distribution is given by

$$f(x|\Lambda_1, \theta_1, \Lambda_2, \theta_2) = e^{\left(-\Lambda_1 e^{-x/\theta_1} - \Lambda_2 e^{-x/\theta_2}\right)} \left(\Lambda_1/\theta_1 \, e^{-x/\theta_1} + \Lambda_2/\theta_2 \, e^{-x/\theta_2}\right), \qquad (9)$$

where $f()$ is the probability density function, x is a mixture of two independent and identically distributed sets of data, $\Lambda_1$ and $\Lambda_2$ are positive relative contributions of the two components, and $\theta_1$ and $\theta_2$ are the exponential random variables of the components.

The Mixed Generalized Extreme Value (Mixed GEV) distribution, also known as the Two-Population Generalized Extreme Value distribution was used by Raynal-Villasenor (2012) [17] to model annual peak flows in Mexico. The probability density function of the Mixed GEV is given by

$$f(x|\mu_1, \sigma_1, \zeta_1, \mu_2, \sigma_2, \zeta_2, \alpha) = \alpha \left[\frac{1}{\sigma} t_1(x)^{\zeta_1+1} e^{-t_1(x)}\right] + (1-\alpha)\left[\frac{1}{\sigma} t_2(x)^{\zeta_2+1} e^{-t_2(x)}\right] \quad (10)$$

and $t(x)$ is

$$t_i(x) = \left(1 + \zeta_i \left(\frac{x-\mu_i}{\sigma_i}\right)\right)^{\frac{-1}{\zeta_i}} \quad \text{if } \zeta_i \neq 0$$

$$t_i(x) = e^{-\frac{x-\mu_i}{\sigma_i}} \quad \text{if } \zeta_i = 0,$$

where $f()$ is the pdf, $i$ is the distribution ($i = 1, 2$), $x$ is a mixture of two independent and identically distributed sets of data, $\mu_1$ and $\mu_2$ are the location parameters for each GEV distribution, $\sigma_1$ and $\sigma_2$ are the scale parameters for each GEV distribution, $\zeta_1$ and $\zeta_2$ are the shape parameters for each distribution, and $\alpha$ is a value [0, 1] defining the weight of the first distribution as a fraction of the whole distribution.

The Mixed Log Pearson III (Mixed LP3) follows the same process as the Mixed GEV. It consists of two LP3 distributions added together with a weighting parameter. This distribution is not unlike the Mixed GEV, but to our knowledge has not been applied in flood frequency analysis before this study.

$$f(x|\tau_1, \alpha_1, \beta_1, \tau_2, \alpha_2, \beta_2, \alpha) = \alpha \left[\frac{\left(\frac{x-\tau_1}{\beta_1}\right)^{\alpha_1-1} e^{\left(-\frac{x-\tau_1}{\beta_1}\right)}}{|\beta_1|\Gamma(\alpha_1)}\right] + (1-\alpha)\left[\frac{\left(\frac{x-\tau_2}{\beta_2}\right)^{\alpha_2-1} e^{\left(-\frac{x-\tau_2}{\beta_2}\right)}}{|\beta_2|\Gamma(\alpha_2)}\right], \quad (11)$$

where $f()$ is the probability density function, $x$ is a mixture of independent and identically distributed sets of data, $\tau_1$ and $\tau_2$ are the location parameters for each LP3 distribution, $\alpha_1$ and $\alpha_2$ are the shape parameters for each LP3 distribution, $\beta_1$ and $\beta_2$ are the scale parameters for each distribution, and $\alpha$ is the relative contribution of the first distribution as a fraction of the whole distribution.

We produce Maximum Likelihood Estimates (MLEs) for each statistical flood frequency distribution using the DEoptim package in R (Table 1) [26]. These MLEs use the annual peak flow, historical flood, and paleoflood bound measurements for Pueblo, Colorado, and account for estimated discharge measurement uncertainty for each data type.

## 2.4 Hydrograph Scaling

Dam safety depends on both the ability to attenuate flood peaks by storing large volumes of water, and to pass high discharges without overtopping. To account for both factors, flood frequency analysis is combined with hydrograph scaling (Fig. 3). Hydrograph scaling creates a flood discharge time series by scaling a representative flood hydrograph to a peak flow. The method creates floods of constant durations, and volumes scaled with the peak flow. The scaled flood hydrographs are routed through the reservoir to determine the reservoir stage. The return period of the peak flow that causes the reservoir to overtop is calculated from the flood frequency curve. This return period is the estimated probability of dam failure by overtopping (Fig. 3). Though a number of flood, earthquake, and static failure scenarios may be relevant to dam safety [13], to maintain the simplicity of this didactic analysis, we consider only dam overtopping.

We sample hydrograph and model uncertainty by considering three representative flood hydrographs and six potential flood frequency models. The three suitably representative flood hydrographs are obtained from three independent sources of information, a physical rainfall-runoff model [14], a regionally representative PMF hydrograph [6], and the flood of record [7]. These hydrographs are time series of roughly durations of half-hour average discharges in units of $m^3s^{-1}$, with the exception of the 1921 flood, the largest flood in the gage record, which is linearly interpolated to half-hour increments from roughly four-hour increments measured during the flood. The hydrographs remain constant in duration while the discharge is scaled linearly such that the maximum half-hour averaged discharge matches the peak flow from the flood frequency curve that the flood hydrograph is being scaled to.

Reservoir stages are calculated by FLROUT, the Bureau of Reclamation flood routing program [13]. FLROUT uses rating curves of reservoir volume-stage and stage-discharge relationships to determine stage and outflows given flood hydrograph inflows for each half-hour time-step. For the dam safety assessment, overtopping peak flows are back-calculated by scaling each of the three flood hydrographs to the smallest peak flow that causes dam overtopping. The return periods for each of the three peak flows are then calculated for each flood frequency model. The method produces three estimates of dam overtopping return period for each flood frequency model. These three estimates are assumed to represent the range of possible dam overtopping return periods for each flood frequency model. In the results, we consider only the Log Pearson Type III and the Mixed GEV distribution results. The results from all six models can be reproduced from the open-source code for this analysis.

## 2.5 Flood hydrograph uncertainty

Uncertainty in the shape of the flood hydrograph is important in the dam overtopping assessment. The duration of the flood and the volume of water delivered over the flood duration determines the flood peak attenuation capacity of the reservoir, and the magnitude of the peak flow through the emergency spillway. The flood hydrographs each last three days but differ in shape, resulting in different total flood volumes for the same peak flow (Fig. 4; Fig. 5).

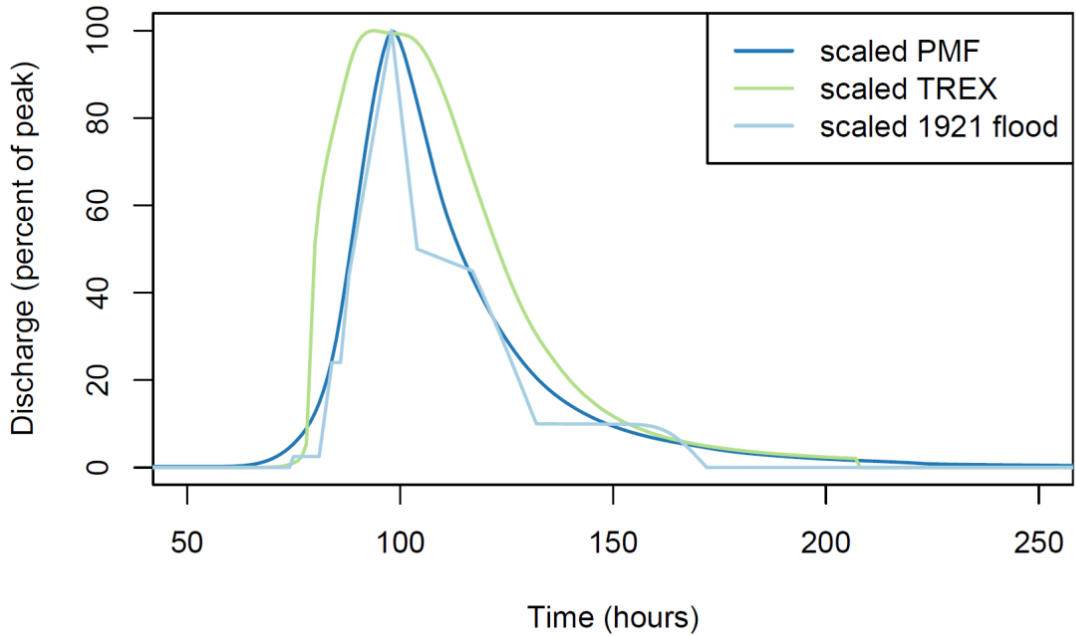

**Fig 4.** Differing shapes of each considered flood hydrograph as a function of time. Each possible flood hydrograph was scaled to the same reference peak flow.

The 1921 "great flood" hydrograph [13] represents the highest peak flow on record. This hydrograph has the lowest volume to peak flow ratio of the three hydrographs (Fig. 4; Fig. 5). For the same peak flow, the reservoir must accommodate a smaller volume of water. The TREX hydrograph is the direct output of the TREX model from stochastic storm transposition, and represents the physically modeled reaction of the system to extreme rainfall [14]. The TREX hydrograph has the largest volume to peak flow ratio of the three hydrographs, and therefore delivers the largest volume for the same peak flow (Fig 4; Fig. 5). The PMF hydrograph is a regionally generalized hydrograph meant to be generally representative of extreme flood hydrographs across multiple watersheds in the hydroclimatic region [6]. The volume to peak flow ratio of the PMF hydrograph is between the 1921 and TREX flood hydrographs and represents the middle estimate (Fig 4; Fig 5).

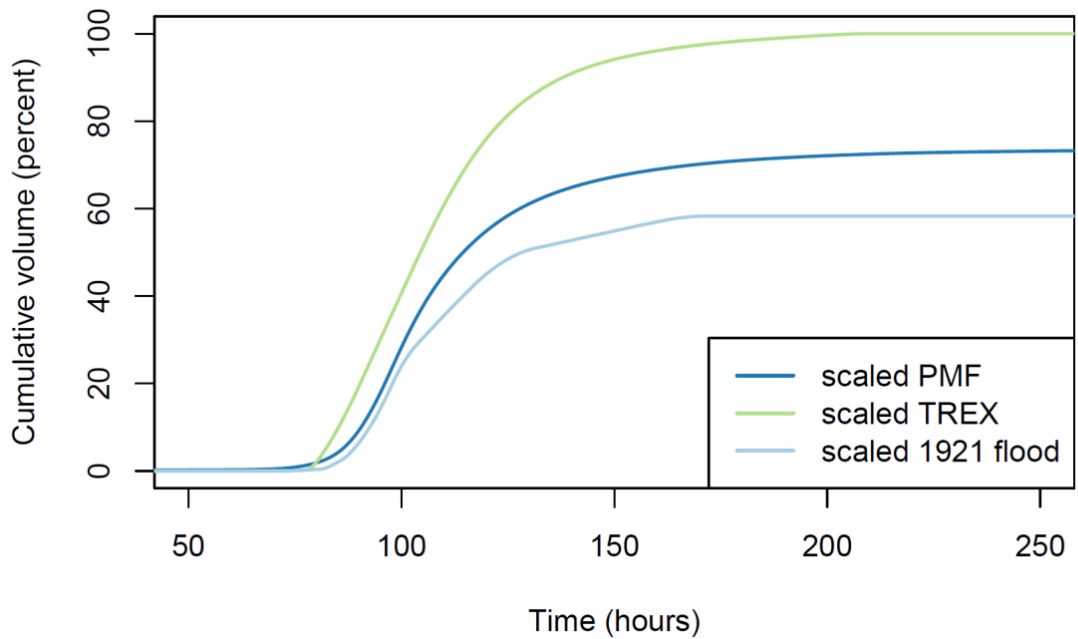

**Fig 5.** The cumulative volume of water delivered over the duration of each considered flood hydrograph as a function of time. Each considered flood hydrograph was scaled to the same reference peak flow.

We route the three potential flood hydrographs through the Pueblo Reservoir to account for hydrograph uncertainty in the dam safety assessment (Fig. 6). The hydrographs are routed with the United States Bureau of Reclamation FLROUT flood routing software [13]. Hydrograph scaling and routing determined the peak flows required for each flood hydrograph to overtop the Pueblo Dam (Fig. 3). We calculate the return periods of the overtopping peak flows for each statistical flood frequency model using the associated flood frequency curve. The return period of the overtopping peak flow determines the return period of dam overtopping and thus the safety of the dam (Fig. 3). Accounting for hydrograph uncertainty changes what would be a deterministic estimate of the relationship between flood exceedance probability and reservoir elevation for each distribution into a cone of uncertainty.

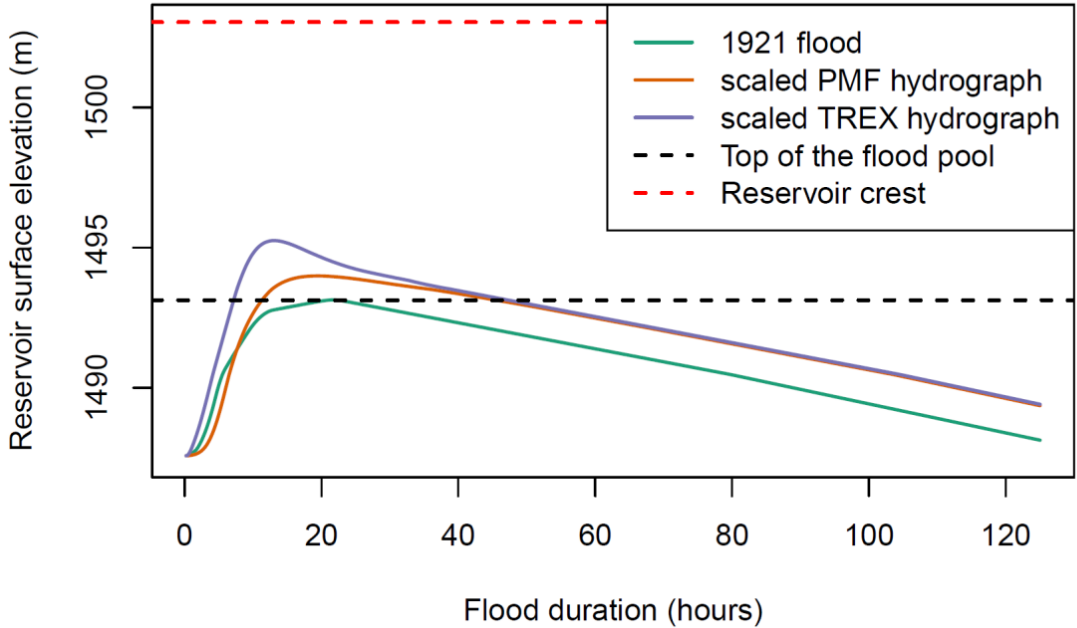

**Fig 6.** Reservoir stage response to each considered flood hydrograph scaled to the 1921 flood peak flow. Differences in cumulative water delivered by each flood hydrograph produce differing reservoir stage responses, and different levels of hazard. The flood pool consists of the extra available storage in the reservoir allotted to flood control. The emergency spillway is at the top of the flood pool. The reservoir crest represents the maximum possible water surface elevation in the reservoir before overtopping. We used FLROUT to route the flood hydrographs.

**2.6 Assessing dam safety**

We classify dam's overtopping-related safety as "meets regulation", "uncertain", or "does not meet regulation" for each distribution based on the return periods of the overtopping floods (Fig. 3). We define the "uncertain" safety interval as return periods between 131,000 and 376,000 years, and the "meets regulations" interval as return periods greater than 376,000 years, while the "does not meet regulations" interval is return periods less than 131,000 years.

**3. Results**

Mixed LP3 and Mixed GEV distributions have complex likelihood spaces due to strong parameter correlations. We assess the performance of the optimization algorithm [26] for maximum likelihood estimation via a simulation study. We compare two independents (each starting from a different random seed) MLEs for each distribution to assess the convergence of the algorithm to the global maximum. For the Mixed LP3, the two MLEs differ by 0% at seven significant figures. For the Mixed GEV, the two MLEs differ by roughly 0.9%. We calculate two more independent MLEs, show that predictions of the 100- and 1,000-year return period floods do not differ largely, and use the highest log-likelihood MLE of the four estimates. We calculate the Akaike Information Criterion (AIC) [27] and Bayesian Information Criterion (BIC) [28] of each model fit to determine goodness-of-fit for each distribution (Table 1). These goodness-of-fit criteria inform the assessment of whether the model choice is subject to overfitting by penalizing models for both higher numbers of parameters and for lower likelihoods.

### 3.1 Best model fit

The Mixed GEV distribution is the best distribution choice according to both AIC and BIC (Table 1). The LP3 is the 4th best fit by the BIC and the 5th best fit by the AIC. Both goodness-of-fit criteria are designed to protect against overfitting, however, the BIC penalizes over parameterization more heavily than the AIC. Our findings support our hypothesis that a mixed distribution model better represents the peak flow data at Pueblo, Colorado (Table 1).

**Table 1** Statistical model types considered in this study, numbers of parameters, and goodness-of-fit criteria. The models are compared with two goodness-of-fit criteria, the Bayesian Information Criterion (BIC), and the Akaike Information Criterion (AIC). Bold font denotes the chosen model for each metric.

| Type | Parameters | BIC | AIC | Statistical model |
|---|---|---|---|---|
| Single | 2 | 1086.763 | 1081.878 | Log Normal (LN2) |
| Single | 3 | 1077.45 | 1070.122 | Log Pearson III (LP3) |
| Single | 3 | 1075.738 | 1068.41 | Generalized Extreme Value (GEV) |
| Mixed | 4 | 1075.557 | 1065.786 | Two Component Extreme Value (TCEV) |
| Mixed | 7 | 1081.016 | 1063.918 | Mixed LP3 |
| **Mixed** | **7** | **1044.062** | **1026.963** | **Mixed GEV** |

### 3.2 Dam safety

The Mixed GEV model predicts higher-discharge events at greater than 104-year return periods relative to the LP3 model, with greater differences at higher return periods. The Mixed GEV predicts higher discharges for events exceeding 104-year return periods, compared to the LP3 model, supporting our second hypothesis that a single-distribution model may under-represent the risk of overtopping at Pueblo Dam. Note, however, that the Mixed GEV model fit still underestimates the size of The Great Flood of 1921 and the paleoflood bound. Nevertheless, the Mixed GEV model underestimates the magnitude of rare events substantially less than the LP3 model (Fig. 7).

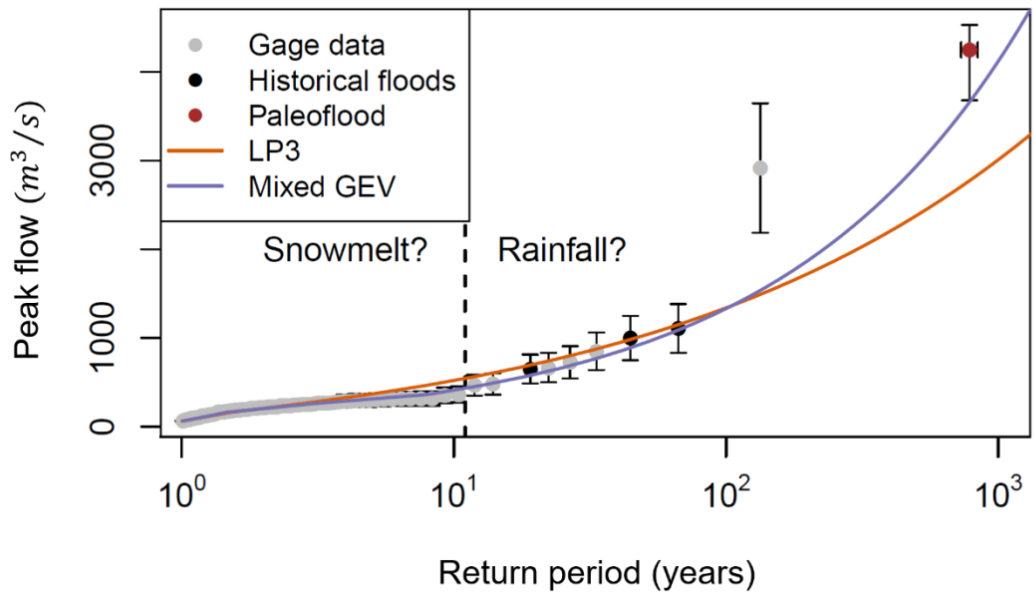

**Fig 7.** Return period plot of the Log Pearson III (LP3) and Mixed Generalized Extreme Value (Mixed GEV) distributions for Pueblo, Colorado.

Switching from the method currently recommended by regulation (LP3) to the statistically better fitting model (Mixed GEV) changes the dam safety classification. Using the three selected hydrographs to represent a reasonable range of outcomes, the LP3 model estimates overtopping return periods of 119,000 to 663,000 years, spanning the range of "meeting", "being uncertain about", and "meeting" the current United States Bureau of Reclamation regulation return periods, 131,000 to 376,000 years (England et al., 2006). However, Mixed GEV model estimates are approximately one order of magnitude shorter, 25,000 to 44,000 years, and do not meet current United States Bureau of Reclamation regulations in any of the three hydrograph scenarios (Fig. 8).

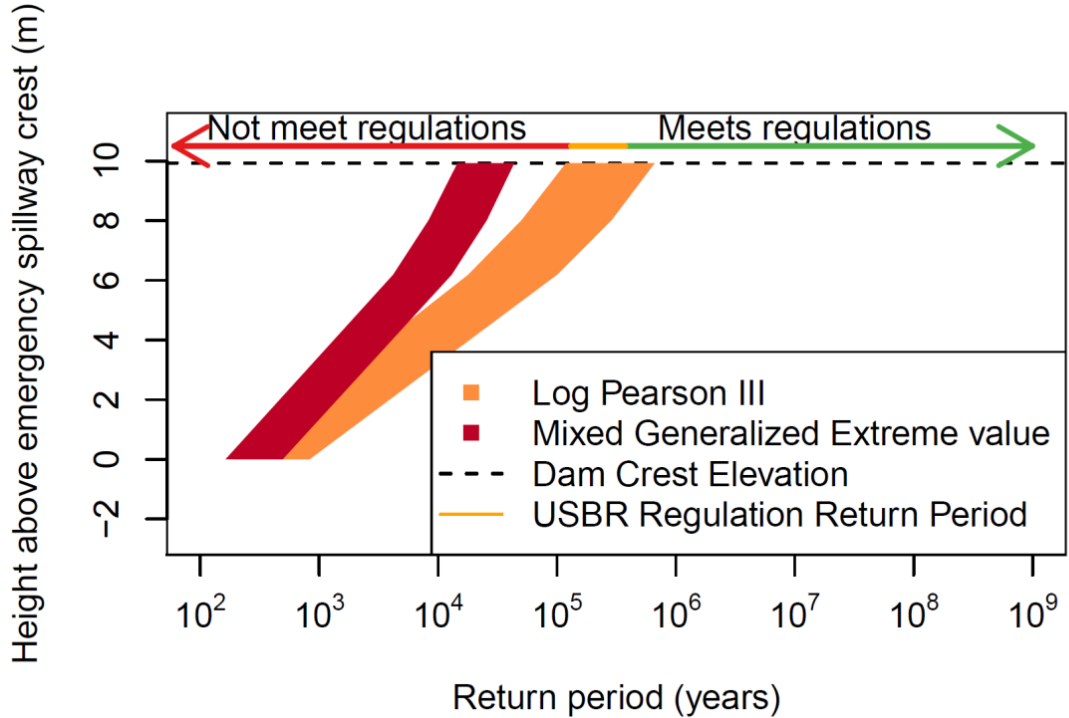

**Fig 8.** Return period plot of reservoir water stage. The bands show uncertainty in reservoir stage for a given peak flow due to flood hydrograph choice.

Flood hydrograph uncertainty presents roughly half an order of magnitude of uncertainty in dam overtopping return periods. This is secondary to model structural uncertainty, which accounts for approximately one order of magnitude in dam overtopping return periods (Fig. 8). Still, this large source of uncertainty is critical for decisions about dam safety (Fig. 8).

## 4. Discussion
### 4.1 Caveats

This case study focuses on method development, model intercomparison, and establishing motivation for further research. This raises several caveats, and the results are not appropriate for actual risk assessments and decision making. For example, we explore mixed distribution statistical flood frequency analysis as a method to account for multiple flood-generating mechanisms while building on existing, computationally efficient approaches to risk characterization. This is only one method of dam safety assessment and is not necessarily the optimal method for the safety assessment of the Pueblo Dam [5]. Rather, the study presents the capabilities of mixed distribution flood frequency models to address mixed annual peak flows without prior knowledge of the physical causes of individual annual peak flows. Additionally, we argue for further consideration of mixed distributions based on the potentially large impacts on assessed flood infrastructure safety.

We focus on Pueblo dam as a case study because the Pueblo Dam is a well-studied area subject to an impactful safety question. The dam safety assessment performed in the study is presented for illustrative purposes rather than actionable results. We present an argument for further investigation of mixed distributions, particularly the Mixed GEV, rather than a complete set of methods for flood frequency analysis and dam safety assessment.

Statistical flood frequency analysis is an important aspect of dam and levee safety assessments, however extrapolation to return periods of 376,000 years from the data available in this analysis, (namely 81 years of gage data, three historical floods, and one approximately 785-year return period paleoflood bound), and indeed the data available in most applications of flood frequency analysis, is subject to substantial statistical and model structural uncertainty. The Bureau of Reclamation recommends physically-based modeling approaches for extreme, high return period events for this reason [5,14]. However, it should also be noted that physically based methods are subject to large uncertainties in watershed and storm behavior, particularly when considering extreme events [14].

We use the robust goodness-of-fit metrics to identify a distribution in an area with two known physical mechanisms of floods and a hypothesized mixed distribution. More testing is necessary before a similar analytical approach is applied without a strong physical justification.

This study uses both AIC and BIC to assess goodness-of-fit and perform model choice. The metrics are derived from different assumptions and calculated with different equations, but are both widely used and accepted as model selection criteria that properly protect against overfitting. Due to the asymptotic nature of these metrics and the small sample size, we strongly recommend caution when these metrics do not clearly agree on a best-fit distribution.

The analysis samples hydrograph uncertainty by repeating the safety assessment using three available flood hydrographs from three independent sources and assuming each is equally likely, and considering no other hydrograph shapes. Further quantification of flood hydrograph variability and uncertainty is an open area of research. These caveats imply that this illustrative analysis is not to be used to assess actual hazards.

## 4.2. Implications

This study illustrates the importance of considering the physical context of statistical models and accounting for mixed distributions where they are present. It also proposes using robust goodness-of-fit metrics to identify mixed distributions with statistical models in flood frequency analysis. Accounting for mixed distributions in risk assessments could result in better statistics and potentially substantial improvements in assessed safety. This has implications for flood risk management developed to satisfy statistically-based design criteria at any locations where multiple physical drivers cause floods.

The LP3 distribution underestimates greater than roughly 105-year return period floods as compared to the Mixed GEV. Underestimations are greater for larger return periods. The 100-year flood calculated by the LP3 fit is a roughly 100-year flood based on the Mixed GEV fit, however, the 500-year flood calculated by the LP3 fit is a roughly 325-year flood based on the Mixed GEV fit, and the 1000-year flood calculated by the LP3 fit is a roughly 525-year flood based on the Mixed GEV fit.

In this analysis, we consider the case of Pueblo, Colorado, where snowmelt causes the majority of peak flows, but in some years extreme rainfall causes peak flows higher than most snowmelt-driven floods. Accounting for the mixed distribution changes the safety assessment of the Pueblo Dam in this didactic example. However, mixed distributions may be present but unaccounted for in many other flood-prone areas. Mixed distributions are already well documented in the Front Range of the Rocky Mountains [10,11]. Additionally, mixed distributions are posited in mountainous Italian watersheds due to multiple rainstorm types [9], and there is the potential for mixed distributions where tropical cyclones dominate the most extreme rainfall events such as the East Coast of the United States [29]. This study has potentially broad implications for flood risk assessments across many regions and flood causes.

We develop methods for identifying mixed distributions and illustrate the importance of accounting for mixed distribution in flood risk analysis, however, we illustrate the numerous difficulties of predicting the size of a roughly 100,000-to-400,000-year return period flood and building to it based on the very limited data available. The best fit statistical method suggests that the dam does not meet current safety regulations for overtopping return periods, while the stochastic storm transposition and rainfall-runoff model approach concludes the dam does meet safety regulations. Additionally, both assessments are associated with large and currently unquantified uncertainties, including climate change.

In the face of extreme statistical and model structural uncertainty, an alternative method may be to adopt adaptive safety standards focused on minimizing loss of life in the case of a disaster. These standards could include improved early warning systems, efficient evacuation plans, or encouraging non-housing development in the potential flood zone. These methods may be a more desirable option for the community given deep uncertainty and finite resources. This study does not address potential tradeoffs between dam safety and increased flood survivability, economic or social, however, these are open questions.

## 5 Conclusions

We use statistical flood frequency models and goodness-of-fit criteria that account for overfitting to assess the presence of a mixed distribution of annual peak flows at the Pueblo Dam. We confirm the presence of mixed distributions of annual peak flows with this method. While caution and further testing are clearly needed, this method may be used to test for the existence of mixed distributions in peak discharges, and more accurately model them, at other locations where current methods can be improved. These methods may help to address shortcomings in the identification and treatment of mixed distributions in flood frequency analysis [8] by contributing to a better understanding of how to identify and deal with mixed distributions.

We find that the mixed GEV is the best model for detecting and representing mixed distribution peak flows. We recommend this model for further investigation. However, the seven-parameter nature of the distribution makes fitting it by maximum likelihoods difficult. Ignoring mixed distributions of peak flows where they are present can lead to serious underestimation of extreme events, as exemplified in this case study. The LP3 distribution fit shows the dam may be safe, uncertain, or unsafe, depending on the choice of flood hydrograph. However, in all flood hydrograph cases, the Mixed GEV predicts the dam does not meet current regulations based on the United States Bureau of Reclamation safety regulations and the loss of life assessment [13].

**Acknowledgments**

We thank the members of the Keller research group for thoughtful discussions about this work. We would also like to thank John F England, Jr. for sharing the TREX hydrographs and for incredibly helpful recommended readings. Special thanks to Randy Miller for reviewing and testing the reproducibility of the analysis code. This study was partially co-supported by the US Department of Energy, Office of Science through the Program on Coupled Human and Earth Systems (PCHES) under DOE Cooperative Agreement No. DE-SC0016162 and the Penn State Center for Climate Risk management. Any opinions, findings and conclusions or recommendations expressed in this material are those of the authors and do not necessarily reflect the views of the funding entities. All coauthors contributed to the interpretation of the results, writing, and revision of the manuscript. The authors do not have any conflicts of interest to declare.

**Code Availability and Disclaimer**

All results, model codes, analysis codes, data, and model outputs used for analysis are freely available from https://github.com/pches/FFA-mixed-distributions and are distributed under the GNU general public license. The datasets, software tools, results, and other resources related to this analysis are provided as-is without warranty of any kind, express or implied. In no event shall the authors or copyright holders be liable for any claim, damages, or other liability in connection with the use of these resources.

**Data Availability**

The United States Geological Survey stream gage annual peak flow record (USGS, 2018a) for this study is available from USGS gage 07099500 at https://nwis.waterdata.usgs.gov/nwis/peak?site_no=07099500&agency_cd=USGS&format=rdb and is automatically downloaded through the freely available open-source code for this study. The historical flood data was compiled from Hafen (1948), Baker and Hafen (1927), Campbell (1922), and Follansbee and Jones (1922) and published in England et al. (2010). The paleoflood data is collected and published in England et al. (2010).